\documentclass[10pt]{article}
\usepackage{multicol}
\usepackage{makeidx}
\usepackage[english]{babel}
\usepackage{graphicx}
\usepackage{amssymb}
\usepackage{epsfig}
\usepackage{hyperref}
\usepackage[english]{babel}
\usepackage{hyperref}
\usepackage[pdftex]{color}
\usepackage{amsmath}
\usepackage{amsfonts}
\begin{document}
\title{\bf On general Lagrangian formulations for arbitrary mixed-symmetric
higher-spin fermionic fields on Minkowski backgrounds}
\author{\textsc{A.  Reshetnyak\thanks{reshet@ispms.tsc.ru, reshet@tspu.edu.ru}  }\\
\it Laboratory of Non-Linear Media Physics, Institute of Strength Physics and  Materials Science,\\ \it  Tomsk, 634021 , Russia.
}
\date{}
\maketitle
\begin{abstract}
The details of unconstrained Lagrangian formulations (being continuation of    earlier developed  research for  Bose particles in [NPB 862 (2012)  270, [arXiv:1110.5044[hep-th]], Phys. of Part. and Nucl. 43 (2012) 689, arXiv:1202.4710 [hep-th]])  are reviewed for Fermi particles propagated on an arbitrary dimensional Minkowski space-time  and described by the unitary irreducible half-integer higher-spin representations of the Poincare group subject to Young tableaux $Y(s_1,...,s_k)$ with $k$ rows. The procedure is based on the construction of the Verma modules and finding auxiliary oscillator realizations for the orthosymplectic $osp(1|2k)$ superalgebra which encodes the second-class operator constraints subsystem in the HS symmetry superalgebra. Applying of an universal BRST-BFV approach permit to reproduce gauge-invariant Lagrangians with reducible gauge symmetries describing the free dynamics of both massless and massive fermionic fields of any spin with appropriate number of gauge and Stukelberg fields. The general construction possesses by the obvious possibility to derive  Lagrangians with only holonomic constraints.
\end{abstract}
\begin{multicols}{2}
\section{Introduction}
The interest to higher-spin (HS) field theory is based on the hopes to reconsider the problems of an unique
description of variety of elementary particles and all known
interactions, in particular,   due to recent success with relating to finding of Higgs boson on LHC\cite{HiggsLHC}. One should remind, that it waits, in addition,  both the proof of
supersymmetry display,  and probably a new insight on origin of Dark Matter
(\cite{LHC}). Due to close interrelation of  HS field theory  to
superstring theory, which operates with an infinite tower of
 HS fields with integer and half-integer spins it  can be viewed  as an method
to study a superstring theory structure. On current state of HS field theory
one may know from the reviews \cite{reviews}, \cite{reviews1}, \cite{reviews2}, \cite{reviews3}. The paper considers
the  results of constructing
 Lagrangian formulations (LFs) for free half-integer   both massless and
 massive
mixed-symmetry spin-tensor HS fields  on flat
$\mathbb{R}^{1,d-1}$-space-time subject to arbitrary Young
tableaux (YT) $Y(n_1,...,n_k)$ for $s_1 = n_1+\frac{1}{2}, \ldots, s_k = n_k+\frac{1}{2}$  in Fronsdal metric-like formalism
on a base of BFV-BRST approach \cite{BFV}, and  precesses  the results
which appear soon in \cite{BuchbinderReshetnyak2} (as continuation of the research for  arbitrary HS fields with integer spin made in
\cite{BuchbinderReshetnyak}, \cite{mixboseResh}).

We know that for higher then  $d=4$ space-time dimensions, there appear,
in addition to totally symmetric irreducible representations of Poincare
or (Anti)-de-Sitter ((A)dS) algebras the mixed-symmetry
representations determined by more than one spin-like parameters
\cite{Labastida},  \cite{metsaevmixirrep}. Whereas  for the former
ones the LFs both for massless and massive free higher-spin fields
is well enough developed \cite{massless Minkowski}, \cite{massless
AdS1}, \cite{massless AdS}, \cite{massive Minkowski},
\cite{massive AdS}, \cite{Francia} as well as on base of BFV-BRST approach, e.g.
in \cite{0001195}--\cite{symferm-ads}, for the latter the problem
of their field-\-theoretic description has not yet solved.
So,
 the main result within the problem  of constrained LF for
 arbitrary massless mixed-symmetry spin-tensor HS fields on a Minkowski space-time
 was obtained in \cite{SkvortsovZinoviev}  in  "frame-like"
 formulation, whereas  in the "metric-like" formulation corresponding
 Lagrangians were  derived in closed manner for only
 reducible Poincare group $ISO(1,d-1)$ representations in \cite{Franciamix}.

We use, first, the conventions for the metric tensor $\eta_{\mu\nu} = diag (+,
-,...,-)$, with Lorentz indices $\mu, \nu = 0,1,...,d-1$, second, the relations $\{\gamma^{\mu}, \gamma^{\nu}\} =
2\eta^{\mu\nu}$,  for Dirac matrices $\gamma^{\mu}$, third, the  notation $\epsilon(A)$, $gh(A)$ for
the respective values of Grassmann parity and ghost number of a
quantity $A$, and denote by $[A,\,B\}$ the supercommutator of
quantities $A, B$, which for theirs definite values of
Grassmann parities is given by $[A\,,B\}$ = $AB -
(-1)^{\epsilon(A)\epsilon(B)}BA$.
%
\section{Half-Integer
HS Symmetry Algebra  for Fermionic fields}\label{Symmalgebra}   A
massless half-integer spin Poincare group irrep  in
$\mathbb{R}^{1,d-1}$  is described by  rank $\sum_{i\geq 1}^k n_i$
spin-tensor field $\Psi_{(\mu^1)_{n_1},(\mu^2)_{n_2},...,(\mu^k)_{n_k}}
\hspace{-0.2em}\equiv \hspace{-0.2em}
\Psi_{\mu^1_1\ldots\mu^1_{n_1},\mu^2_1\ldots\mu^2_{n_2},...,}$
${}_{ \mu^k_1\ldots \mu^k_{n_k}{}A}(x)$
  with generalized spin
 $\mathbf{s} = (n_1+\frac{1}{2}, n_2+\frac{1}{2},
 ... , n_k+\frac{1}{2})$, ($n_1 \geq n_2\geq ... \geq n_k>0, k
\leq [(d-1)/2])$ subject to a YT with $k$ rows of  lengths $n_1, n_2,
..., n_k$ and suppressed Dirac index $A$,
\begin{equation}\label{Young k}
\Psi_{(\mu^1)_{n_1},(\mu^2)_{n_2},...,(\mu^k)_{n_k}}
\leftrightarrow
\begin{array}{|c|c  c|c|c|c| c|}\hline
  \!\!\mu^1_1\!\! & \!\!\cdot  \!\!  & \!\!\cdot\!\!  & \!\!\cdot\!\!  &\!\! \cdot\!\! &
  \!\!\cdot   \!\! & \!\!\mu^1_{n_1}\!\!\! \\
   \hline
     \!\!\mu^2_1 \!\!
   &\!\! \cdot \!\!&\!\! \cdot  \!\! &  \!\cdot\! & \!\!\mu^2_{n_2}\!\!   \\
  \cline{1-5} \cdots   \\
   \cline{1-4}
     \!\mu^k_1 \!& \!\cdot \!& \!\cdot\! &\! \mu^k_{n_k}\! \!\!  \\
   \cline{1-4}
\end{array}\ ,
\end{equation}
The spin-tensor is symmetric with respect to the permutations of each
type of indices $\mu^i$
  and
obeys to the Dirac (\ref{Eq-1}), gamma-traceless
(\ref{Eq-2}) and mixed-symmetry equations (\ref{Eq-3}) [for
$i,j=1,...,k;\, l_i,m_i=1,...,n_i$]:
\begin{eqnarray}
&&
\imath\gamma^{\mu}\partial_{\mu}\Psi_{(\mu^1)_{n_1},(\mu^2)_{n_2},...,(\mu^k)_{n_k}}
=0,   \label{Eq-1}
\\
&& \gamma^{\mu^i_{l_i}}\Psi_{
(\mu^1)_{n_1},(\mu^2)_{n_2},...,(\mu^k)_{n_k}} =0,  \label{Eq-2}\\
&& \Psi_{
(\mu^1)_{n_1},...,\{(\mu^i)_{n_i}\underbrace{,...,\mu^j_{1}...}\mu^j_{l_j}\}...\mu^j_{n_j},...(\mu^k)_{n_k}}=0,
 \label{Eq-3}
\end{eqnarray}
for $i<j,\ 1\leq l_j\leq n_j$ and where the  bracket below denote that the indices  in it do not
include in  symmetrization.

Joint description of all half-integer spin $ISO(1,d-1)$ group irreps can be standardly
reformulated  with an auxiliary Fock space
$\mathcal{H}$, generated by $k$ pairs of bosonic creation
$a^i_{\mu^i}(x)$ and annihilation $a^{j+}_{\nu^j}(x)$ operators (in symmetric basis),
$i,j =1,...,k, \mu^i,\nu^j =0,1...,d-1$: $[a^i_{\mu^i},
a_{\nu^j}^{j+}]=-\eta_{\mu^i\nu^j}\delta^{ij}$
 and a set of constraints for an arbitrary string-like  (so called \emph{basic}) vector
$|\Psi\rangle \in \mathcal{H}$, being as well Dirac spinor,
\begin{eqnarray}
\label{PhysState}  \hspace{-2ex}&& \hspace{-2ex} |\Psi\rangle  =
\sum_{n_1=0}^{\infty}\sum_{n_2=0}^{n_1}\cdots\sum_{n_k=0}^{n_{k-1}}
\Psi_{(\mu^1)_{n_1},(\mu^2)_{n_2},...,(\mu^k)_{n_k}}\nonumber \\
&&\quad \times
\prod_{i=1}^k\prod_{l_i=1}^{n_i} a^{+\mu^i_{l_i}}_i|0\rangle,\\
\label{l0} \hspace{-2ex}&&\hspace{-2ex}  \bigl(\tilde{t}_0, \tilde{t}^i,  t^{i_1j_1}
\bigr)|\Psi\rangle = 0,\texttt{ for }i\leq j;\, i_1 < j_1,
\end{eqnarray}
where $\bigl(\tilde{t}_0, \tilde{t}^i,  t^{i_1j_1}
\bigr) =  \bigl( -i\gamma^{\mu}\partial_\mu, \gamma^{\mu}a^i_\mu,
a^{i_1+}_\mu a^{j_1\mu}\bigr)$.

 The set of $(\frac{1}{2}k(k+1)+1)$
primary constraints (\ref{l0}), $\{o_\alpha\}$ = $\bigl\{\tilde{t}_0,
\tilde{t}^i, t^{i_1j_1} \bigr\}$, with additional
condition, $g_0^i|\Psi\rangle =(n_i+\frac{d}{2}) |\Psi\rangle$ for
 number particles operators, $g_0^i = - a^{i+}_\mu  a^{\mu{}i} +
 \frac{d}{2}$, are equivalent to Eqs.
(\ref{Eq-1})--(\ref{Eq-3}) for given spin
$\mathbf{s}$.

The fermionic nature of equations
(\ref{Eq-1}), (\ref{Eq-2}) and  the bosonic one of
the primary constraint operators $\tilde{t}_0, \tilde{t}^i$  with respect to the standard
Lorentz-like Grassmann parity,
$\epsilon(\tilde{t}_0) = \epsilon(\tilde{t}^i)= 0$ are in contradiction and resolve the problem of, $(\tilde{t}^i)^2 = \frac{1}{2}{\gamma}^\mu{\gamma}^\nu a^i_\mu a^i_\nu+a^i_\nu a^i_\mu = a^i_\mu a^i_\mu = 2 l^{ii}$, with new "traceless"  operator, we equivalently transform  above operators into fermionic ones.
Following to Ref. \cite{symferm-flat}, \cite{symferm-ads} we
introduce a set of $(d+1)$ Grassmann-odd gamma-matrix-like objects
$\tilde{\gamma}^\mu$, $\tilde{\gamma}$, subject to
\begin{eqnarray}
\{\tilde{\gamma}^\mu,\tilde{\gamma}^\nu\} = 2\eta^{\mu\nu}, \qquad
\{\tilde{\gamma}^\mu,\tilde{\gamma}\}=0, \qquad
\tilde{\gamma}^2=-1, \label{tgammas}
\end{eqnarray}
and related  to the conventional gamma-matrices as: $
\gamma^{\mu} = \tilde{\gamma}^{\mu} \tilde{\gamma}$.

Therefore, the odd constraints,
\begin{eqnarray}
\label{t0ti} {t}_0 = -\imath\tilde{\gamma}^\mu \partial_\mu\,,
\qquad\qquad {t}^i =  \tilde{\gamma}^\mu a^i_\mu ,
\end{eqnarray}
%
related to the operators (\ref{l0}) as:
$\left(t_0, t^i\right) = \tilde{\gamma}\left(\tilde{t}_0,
\tilde{t}^i\right)$, solve the problem above.

The finding of Lagrangian as, $\mathcal{L} \sim \langle \Psi Q \Psi\rangle$,  implies  the Hermiticity  of BFV-BRST
operator $Q$, $Q = C^\alpha o_\alpha + \ldots$,   that means the
extension of the set $\{o_\alpha\}$ up to one of $\{o_I\} =
\{o_\alpha, l_i, l_{ij}; o^+_\alpha, l^+_i, l^+_{ij}; g_0^i\}$,  for divergent and gradient operators $(l_i, l^+_i) = -  \imath ( a^\mu_i\partial_\mu, a^{\mu+}_i\partial_\mu)$ and for   $l^+_{ij} = \frac{1}{2}a^{\mu+}_ia^{\mu+}_j$, $i\le j$, which is closed with respect to supercommutator multiplication $[\ ,\ \}$ and
hermitian conjugation related to odd scalar product on
$\mathcal{H}$,
\begin{eqnarray}
\label{sproduct} \hspace{-1ex}&\hspace{-1ex}&\hspace{-1ex} \langle\tilde{\Phi}|\Psi\rangle  =   \int
d^dx\sum_{(n)_k=0}^{\infty,(n)_{k-1}} \sum_{(p)_k=0}^{\infty,(p)_{k-1}}
\langle 0|\prod_{j=1;m_j=1}^{k;p_j}
a^{\nu^j_{m_j}}_j\nonumber\\
\hspace{-1ex}&\hspace{-1ex}&\hspace{-1ex} \times\Phi^+_{(\nu^1)_{p_1},(\nu^2)_{p_2},...,(\nu^k)_{p_k}} \tilde{\gamma}_0
\Psi_{(\mu^1)_{n_1},(\mu^2)_{n_2},...,(\mu^k)_{n_k}}\nonumber\\
\hspace{-1ex}&\hspace{-1ex}&\hspace{-1ex} \qquad \times
\prod_{i=1; l_i=1}^{k;n_i} a^{+\mu^i_{l_i}}_i|0\rangle ,
\end{eqnarray}
for $\sum_{(n)_k=0}^{\infty,(n)_{k-1}} \equiv \sum_{n_1=0}^{\infty}\sum_{n_2=0}^{n_1}...\sum_{n_{k}=0}^{n_{k-1}}$, $n_i, p_j \in \mathbb{N}_0$. Operators
$o_I$ satisfy to the Lie-superalgebra commutation relations,
    $[o_I,\ o_J\}= f^K_{IJ}o_K$, for structure constants
    $f^K_{IJ}= - (-1)^{\epsilon(o_I)\epsilon(o_J)}f^K_{JI}$, to be determined by anticommutators,
    \begin{eqnarray}
    \hspace{-1ex}&\hspace{-1ex}&\hspace{-1ex}  \{t_0, t_0\}\ =\ -2 l_0, \quad \{t_0, t_i\}= 2l_i, \quad \{t_{i}, t_j\}\ =\ 4 l_{ij},
    \nonumber \\
\hspace{-1ex}&\hspace{-1ex}&\hspace{-1ex}       \quad \{t_{i}, t_j^+\} \ =\ 2\bigl(-g_0^{i}\delta_{ij} +
   t_{ji}\theta^{ij} +
   t_{ij}^+\theta^{ji} \bigr), \label{oddmult}
    \end{eqnarray}
 (with Heaviside $\theta$-symbol $\theta^{ij}$)   and  from the
multiplication table~\ref{table in} with only commutators.

The products $B^{i_2j_2}_{i_1j_1}, A^{i_2j_2, i_1j_1},
F^{i_1j_1,i}, L^{i_2j_2,i_1j_1}$ in the table~\ref{table in} are
given by the  relations,
\begin{eqnarray}
  &&\hspace{-1em}{}B^{i_2j_2}{}_{i_1j_1}\ =\
  (g_0^{i_2}-g_0^{j_2})\delta^{i_2}_{i_1}\delta^{j_2}_{j_1} +
  (t_{j_1}{}^{j_2}\theta^{j_2}{}_{j_1} \label{Bijkl}\\
  && + t^{j_2}{}^+_{j_1}\theta_{
  j_1}{}^{j_2})\delta^{i_2}_{i_1}
  -(t^+_{i_1}{}^{i_2}\theta^{i_2}{}_{i_1} + t^{i_2}{}_{i_1}\theta_{i_1}{}^{
  i_2})
  \delta^{j_2}_{j_1}
\,,\nonumber\\
  &&\hspace{-1em}A^{i_2j_2, i_1j_1} =  t^{i_1j_2}\delta^{i_2j_1}-
  t^{i_2j_1}\delta^{i_1j_2}  ,\ F^{i_2j_2,i}  =
   t^{[i_2j_2}\delta^{j_2]i},\label{Fijk} \nonumber
    \end{eqnarray}
\end{multicols}
{\begin{table}[t] {{\footnotesize
\begin{center}
\begin{tabular}{||c||c|c|c|c|c|c|c||c||}\hline\hline
$\hspace{-0.2em}[\; \downarrow, \rightarrow
]\hspace{-0.5em}$\hspace{-0.7em}&
 $t^{i_1j_1}$ & $t^+_{i_1j_1}$ &
$l_0$ & $l^i$ &$l^{i{}+}$ & $l^{i_1j_1}$ &$l^{i_1j_1{}+}$ &
$g^i_0$ \\ \hline\hline $t_0$
    & $0$ & $0$
   & $0$&\hspace{-0.3em}
    $0$\hspace{-0.5em} &
    \hspace{-0.3em}
    $0$\hspace{-0.3em}
    &\hspace{-0.7em} $0$ \hspace{-1.2em}& \hspace{-1.2em}$
    0$\hspace{-1.2em}& $0$ \\
\hline $t^{i_2}$
    & $-t^{j_1}\delta^{i_2i_1}$ & $-t_{i_1}\delta^{i_2}{}_{j_1}$
   & $0$&\hspace{-0.3em}
    $\hspace{-0.2em}0$\hspace{-0.5em} &
    \hspace{-0.3em}
    $-t_0\delta^{i_2i}$\hspace{-0.3em}
    &\hspace{-0.7em} $0$ \hspace{-1.2em}& \hspace{-1.2em}$
    -\frac{1}{2}t^{\{i_1+}\delta^{j_1\}i_2}\hspace{-0.9em}$\hspace{-1.2em}& $t^{i_2}\delta^{i_2i}$ \\
\hline$t^{i_2+}$
    & $t^{i_1+}\delta^{i_2j_1}$ & $t^+_{j_1}\delta_{i_1}{}^{i_2}$
   & $0$&\hspace{-0.3em}
    $\hspace{-0.2em}t_0\delta^{i_2i}$\hspace{-0.5em} &
    \hspace{-0.3em}
    $0$\hspace{-0.3em}
    &\hspace{-0.7em} $\hspace{-0.7em}\frac{1}{2}t^{\{i_1}\delta^{j_1\}i_2}
    \hspace{-0.9em}$ \hspace{-1.2em}& \hspace{-1.2em}$0\hspace{-0.9em}$\hspace{-1.2em}& $-t^{i_2+}\delta^{i_2i}$ \\
\hline\hline $t^{i_2j_2}$
    & $A^{i_2j_2, i_1j_1}$ & $B^{i_2j_2}{}_{i_1j_1}$
   & $0$&\hspace{-0.3em}
    $\hspace{-0.2em}l^{j_2}\delta^{i_2i}$\hspace{-0.5em} &
    \hspace{-0.3em}
    $-l^{i_2+}\delta^{j_2 i}$\hspace{-0.3em}
    &\hspace{-0.7em} $\hspace{-0.7em}l^{\{j_1j_2}\delta^{i_1\}i_2}
    \hspace{-0.9em}$ \hspace{-1.2em}& \hspace{-1.2em}$
    -l^{i_2\{i_1+}\delta^{j_1\}j_2}\hspace{-0.9em}$\hspace{-1.2em}& $F^{i_2j_2,i}$ \\
\hline $t^+_{i_2j_2}$
    & $-B^{i_1j_1}{}_{i_2j_2}$ & $A^+_{i_1j_1, i_2j_2}$
&$0$   & \hspace{-0.3em}
    $\hspace{-0.2em} l_{i_2}\delta^{i}_{j_2}$\hspace{-0.5em} &
    \hspace{-0.3em}
    $-l^+_{j_2}\delta^{i}_{i_2}$\hspace{-0.3em}
    & $l_{i_2}{}^{\{j_1}\delta^{i_1\}}_{j_2}$ & $-l_{j_2}{}^{\{j_1+}
    \delta^{i_1\}}_{i_2}$ & $-F_{i_2j_2}{}^{i+}$\\
\hline $l_0$
    & $0$ & $0$
& $0$   &
    $0$\hspace{-0.5em} & \hspace{-0.3em}
    $0$\hspace{-0.3em}
    & $0$ & $0$ & $0$ \\
\hline $l^j$
   & \hspace{-0.5em}$- l^{j_1}\delta^{i_1j}$ \hspace{-0.5em} &
   \hspace{-0.5em}$
   -l_{i_1}\delta_{j_1}^{j}$ \hspace{-0.9em}  & \hspace{-0.3em}$0$ \hspace{-0.3em} & $0$&
   \hspace{-0.3em}
   $l_0\delta^{ji}$\hspace{-0.3em}
    & $0$ & \hspace{-0.5em}$- \textstyle\frac{1}{2}l^{\{i_1+}\delta^{j_1\}j}$
    \hspace{-0.9em}&$l^j\delta^{ij}$  \\
\hline $l^{j+}$ & \hspace{-0.5em}$l^{i_1+}
   \delta^{j_1j}$\hspace{-0.7em} & \hspace{-0.7em}
   $l_{j_1}^+\delta_{i_1}^{j}$ \hspace{-1.0em} &
   $0$&\hspace{-0.3em}
      \hspace{-0.3em}
   $-l_0\delta^{ji}$\hspace{-0.3em}
    \hspace{-0.3em}
   &\hspace{-0.5em} $0$\hspace{-0.5em}
    &\hspace{-0.7em} $ \textstyle\frac{1}{2}l^{\{i_1}\delta^{j_1\}j}
    $\hspace{-0.7em} & $0$ &$-l^{j+}\delta^{ij}$  \\
\hline $l^{i_2j_2}$
    & \hspace{-0.3em}$\hspace{-0.4em}-l^{j_1\{j_2}\delta^{i_2\}i_1}\hspace{-0.5em}$
    \hspace{-0.5em} &\hspace{-0.5em} $\hspace{-0.4em}
    -l_{i_1}{}^{\{i_2+}\delta^{j_2\}}_{j_1}\hspace{-0.3em}$\hspace{-0.3em}
   & $0$&\hspace{-0.3em}
    $0$\hspace{-0.5em} & \hspace{-0.3em}
    $ \hspace{-0.7em}-\textstyle\frac{1}{2}l^{\{i_2}\delta^{j_2\}i}
    \hspace{-0.5em}$\hspace{-0.3em}
    & $0$ & \hspace{-0.7em}$\hspace{-0.3em}
L^{i_2j_2,i_1j_1}\hspace{-0.3em}$\hspace{-0.7em}& $\hspace{-0.7em}  l^{i\{i_2}\delta^{j_2\}i}\hspace{-0.7em}$\hspace{-0.7em} \\
\hline $l^{i_2j_2+}$
    & $ l^{i_1 \{i_2+}\delta^{j_2\}j_1}$ & $ l_{j_1}{}^{\{j_2+}
    \delta^{i_2\}}_{i_1}$
   & $0$&\hspace{-0.3em}
    $\hspace{-0.2em} \textstyle\frac{1}{2}l^{\{i_2+}\delta^{ij_2\}}$\hspace{-0.5em} & \hspace{-0.3em}
    $0$\hspace{-0.3em}
    & $-L^{i_1j_1,i_2j_2}$ & $0$ &$\hspace{-0.5em}  -l^{i\{i_2+}\delta^{j_2\}i}\hspace{-0.3em}$\hspace{-0.2em} \\
\hline\hline $g^j_0$
    & $-F^{i_1j_1,j}$ & $F_{i_1j_1}{}^{j+}$
   &$0$& \hspace{-0.3em}
    $\hspace{-0.2em}-l^i\delta^{ij}$\hspace{-0.5em} & \hspace{-0.3em}
    $l^{i+}\delta^{ij}$\hspace{-0.3em}
    & \hspace{-0.7em}$\hspace{-0.7em}  -l^{j\{i_1}\delta^{j_1\}j}\hspace{-0.7em}$\hspace{-0.7em} & $ l^{j\{i_1+}\delta^{j_1\}j}$&$0$ \\
   \hline\hline
\end{tabular}
\end{center}}} \vspace{-2ex}\caption{even-even and odd-even parts of HS symmetry  superalgebra  $\mathcal{A}^f(Y(k),
\mathbb{R}^{1,d-1})$.\label{table in} }\end{table}
\vspace{-4ex}\begin{multicols}{2}
\begin{eqnarray}
  &&\hspace{-1em}L^{i_2j_2,i_1j_1} =   \textstyle\frac{1}{4}\hspace{-0.15em}
  \bigl\{\delta^{i_2i_1}
\delta^{j_2j_1}\hspace{-0.15em}\bigl[\hspace{-0.15em}
2g_0^{i_2}\delta^{i_2j_2} \hspace{-0.15em}+\hspace{-0.15em}
g_0^{i_2}\hspace{-0.15em} +\hspace{-0.15em}
g_0^{j_2}\hspace{-0.15em}\bigr] \hspace{-0.15em} \nonumber \\
&& -
\hspace{-0.15em}\bigl(\hspace{-0.15em}\delta^{j_2\{i_1}\hspace{-0.15em}
\bigl[\hspace{-0.15em}t^{j_1\}i_2}\theta^{i_2j_1\}}
\hspace{-0.15em} +\hspace{-0.15em}t^{i_2j_1\}+}\theta^{j_1\}i_2}\hspace{-0.15em}\bigr]\hspace{-0.15em}
\hspace{-0.15em}+ \hspace{-0.15em}(j_2\leftrightarrow
i_2)\hspace{-0.15em}\bigr)\hspace{-0.25em}\bigr\}.  \nonumber \label{Lklij}
\end{eqnarray}
We call the superalgebra of the operators $o_I$
as \emph{half-integer higher-spin symmetry algebra in Minkowski
space with a YT having $k$ rows} and denote it as
$\mathcal{A}^f(Y(k), \mathbb{R}^{1,d-1})$. It appears as the superextension of
\emph{integer higher-spin symmetry algebra} $\mathcal{A}(Y(k), \mathbb{R}^{1,d-1})$ introduced in
 \cite{BuchbinderReshetnyak} for bosonic fields.

Hamiltonian
analysis of the topological dynamical system of
 the operators $\{o_I\}$ permits to  classify   $2k(k+1)$ operators $\{o_a\}=\{t^i, l^{ij},t^{i_1j_1}, t^+_i, l^+_{ij},t^+_{i_1j_1}
 \}$ as second-class and
   $2(k+1)$ ones  $\{t_0, l_0, l^{i}, l^+_{j} \}$ as first-class constraints
  subsystems whereas $k$ elements $g_0^i$ form  supermatrix $\Delta_{ab}(g_0^i)$ in $[o_a,o_b\} \sim
  \Delta_{ab}$. The subsystem of the second-class constraints $\{o_a\}$ together
with $\{g_0^i\}$ forms the subalgebra in $\mathcal{A}(Y(k),
\mathbb{R}^{1,d-1})$ to be isomorphic, due to Howe duality \cite{Howe1}, to  orthosymplectic $osp(1|2k)$
algebra (the details, see in \cite{BuchbinderReshetnyak2}).

The construction of the HS symmetry superalgebra $\mathcal{A}^f(Y(k), \mathbb{R}^{1,d-1})$,  can  not permit to
construct BRST operator $Q$ with respect to its elements $o_I$
  due to second-class
constraints $\{o_a\}$ presence in it. Therefore we should to convert
orthosymplectic algebra $osp(1|2k)$ of $\{o_a,g_0^i\}$  into enlarged set
of operators $O_I$ with only first-class constraints.

\section{Scalar Oscillator realization for $osp(1|2k)$
}\label{Vermamodule}

We consider  an additive conversion procedure developed within BRST
method, (see e.g. \cite{BurdikPashnev}),   implying  the
enlarging of $o_I$ to $O_I = o_I + o'_I$, with additional parts
$o'_I$ supercommuting with all $o_I$ and determined on a new Fock space $\mathcal{H}'$. Now, the
elements $O_I$ are given on $\mathcal{H}\otimes \mathcal{H}'$ so
that a condition for $O_I$, $[O_I,\ O_J] \sim O_K$, leads to the
same algebraic relations for $O_I$ and $o'_I$ as those for $o_I$.

Not going into details  of Verma module (special representation
space \cite{Dixmier}) construction for the superalgebra
$osp(1|2k)$ of new operators $o'_I$ considered in \cite{BuchbinderReshetnyak2} and for the case of its $sp(k|2k)$ subalgebra in
\cite{BuchbinderReshetnyak}, we  present here theirs explicit
oscillator form in terms of new $2k(k+1)$ creation and annihilation
operators $(B^+_d,B^d;)$ = $(f^+_i,b^{+}_{ij}, d^+_{rs}; f_i, b_{ij},
d_{rs})$, $i,j,r,s =1,\ldots, k; i\leq j; r<s$ as follows (for
$k_0\equiv l$)
\begin{eqnarray}
\hspace{-1em}&&\hspace{-1em}g_0^{\prime i}\ = \
f^+_if_i + \sum_{l\leq m}
 b_{lm}^+b_{lm}(\delta^{il}+\delta^{im}) \nonumber\\
 \hspace{-1em}&&\hspace{0em} + \sum_{r< s}d^+_{rs}d_{rs}(\delta^{is}-
 \delta^{ir}) +h^i
 \,,\label{g'0iF}\\
 \hspace{-1em}&&\hspace{-1em}l^{\prime+}_{ij} \ = \
b_{ij}^+\,,\quad t^{\prime  +}_i  =  f^+_i + 2b_{ii}^+f_
 +4\sum_{l=1}^{i-1}b_{li}^+f_l\,,\label{lij+F} \\
\hspace{-1em}&&\hspace{-1em} t^{\prime+}_{lm}   = d^+_{lm} -
\sum_{n=1}^{l-1}d_{nl}d^+_{nm}
   - \sum_{n=1}^{k}(1+\delta_{nl})b^+_{nm}b_{ln}\,,
 \label{t'+lmtext}
  \end{eqnarray}
  \begin{eqnarray}
\vspace{-1ex}
\hspace{-2em}&&\hspace{-1em}t^{\prime }_{lm} = -
\sum_{n=1}^{l-1}d^+_{nl}d_{nm}-\sum_{n=1}^{k}(1+\delta_{nm})b^+_{nl}
b_{nm}\nonumber\\
 \hspace{-2em}&&\hspace{0em} +
\sum_{p=0}^{m-l-1}\hspace{-0.2em}\sum_{k_1=l+1}^{m-1}... \hspace{-0.2em}\sum_{k_p=l+p}^{m-1}
 \hspace{-0.2em}C^{k_{p}m}(d^+,d)\prod_{j=1}^pd_{k_{j-1}k_{j}}
\nonumber\\
 \hspace{-2em}&&\hspace{0em}
    + \bigl[4\sum_{n=r+1}^{s-1}b^+_{rn}f_n +(2b^+_{rr}f_r-{f}^+_{r})\bigr]f_{s}
 \label{t'lmF}.\end{eqnarray}
Note, first, that $B_c, B^+_d$ satisfy to the standard commutation
relations, $[B_c, B^+_d\}= \delta_{cd}$, for instance, $\{f_i, f^+_j\} = \delta_{ij}$, for odd $f_i, f^+_j$.   Second, the arbitrary
parameters $h^i$ in (\ref{g'0iF}) need to reproduce correct LF for
HS field with given spin $\mathbf{s}$, whereas the form of the
rest elements $t^{\prime }_{i}, l^{\prime }_{ij}$, for $i\leq j$, to be expressed
by means of the operators $C^{lm}(d^+,d), l<m$, as well as the property of Hermiticity
 for them may be found in \cite{BuchbinderReshetnyak}, \cite{BuchbinderReshetnyak2}\footnote{The
 case of  massive   HS fields whose
system of $2$nd-class constraints contains additionally to
elements of $osp(1|2k)$ superalgebra  the constraints  of isometry
subalgebra of  Minkowski space  $t_0, l^i, l^+_i, l_0$ may be treated
by dimensional reduction of the algebra
$\mathcal{A}^f(Y(k),\mathbb{R}^{1,d})$ for massless HS fields to one
$\mathcal{A}^f(Y(k),\mathbb{R}^{1,d-1})$ for massive HS fields, (see
 \cite{BuchbinderReshetnyak2}). Now, the Dirac equation in
(\ref{Eq-1}) is changed on massive equation corresponding to
the constraint $t_0$ ($t_0=-\imath\tilde{\gamma}^\mu \partial_\mu +\tilde{\gamma}m$)  acting
on the same  \emph{basic} vector $|\Psi\rangle$
(\ref{PhysState}).}.


\section{BRST-BFV operator and Lagrangian formulations}
\label{BRSToperator} Due to  algebra of $O_I$ under
consideration is a Lie superalgebra
$\mathcal{A}^f(Y(k),\mathbb{R}^{1,d-1})$ the BFV-BRST operator $Q'$
may be constructed in the standard way as
\begin{equation}\label{generalQ'}
    Q'  = {O}_I\mathcal{C}^I + \textstyle\frac{1}{2}
    \mathcal{C}^I\mathcal{C}^Jf^K_{JI}\mathcal{P}_K(-1)^{\epsilon({O}_K) + \epsilon({O}_I)}
\end{equation}
with the constants $f^K_{JI}$ from the table~\ref{table in},
constraints $O_I = (T_0, T^+_i$, $T_i$;  $L_0, L^+_i$, $L_i$, $L_{ij}, L^+_{ij},
T_{rs}$, $T^+_{rs}, G_0^i)$, fermionic [bosonic]  ghost fields and
conjugated to them momenta $(C^I, \mathcal{P}_I)$  =
$\bigl((\eta_0, {\cal{}P}_0); (\eta^i, {\cal{}P}^+_i)$;
$(\eta^+_i$, ${\cal{}P}_j)$; $(\eta^{ij}, {\cal{}P}^+_{ij})$;
$(\eta^+_{ij}, {\cal{}P}_{ij}); (\vartheta_{rs},\lambda^+_{rs})$;
$(\vartheta^+_{rs}, \lambda_{rs}); (\eta^i_{G},
{\cal{}P}_{G})\bigr)$, $[(q_0, p_0), (q_i^+, p_i), (q_i, p_i^+)]$  with the properties
\begin{eqnarray}
 \hspace{-1em}&&\hspace{-1em}
    (\eta^{ij},\vartheta_{rs})= (\eta^{ji} , \vartheta_{rs}\theta^{sr})
,\ \{\vartheta_{rs},\lambda^+_{tu}\}= \delta_{rt}\delta_{su},\nonumber\\
\hspace{-1em}&&\hspace{-1em}\label{propgho} \{\eta_{lm},{\cal{}P}_{ij}^+\}= \delta_{li}\delta_{jm}, \ \{{\cal{}P}_j, \eta_i^+\}= [q_i, p^{+}_j] = \delta_{ij}
\end{eqnarray}
 and non-vanishing (anti)commutators $\{\eta_0,{\cal{}P}_0\}= [q_0, p_0]= \imath$,
$\{\eta^i_{\mathcal{G}}, {\cal{}P}^j_{\mathcal{G}}\}$
 = $\imath\delta^{ij}$ for zero-mode ghosts\footnote{The ghosts possess the
 standard  ghost number distribution,
$gh(\mathcal{C}^I)$ = $ - gh(\mathcal{P}_I)$ = $1$
$\Longrightarrow$  $gh({Q}')$ = $1$.}.

To construct LF for fermionic HS fields in a $\mathbb{R}^{1,d-1}$
 space-time we partially follow the algorithm of
\cite{mixfermiflat}, being a particular
case of our construction for $n_3 = 0$. First, we
extract the dependence of  $Q'$ (\ref{generalQ'}) on the ghosts
$\eta^i_{G}, {\cal{}P}^i_{G}$, to obtain generalized spin operator $\sigma^i$ and the BRST operator $Q$
only for the system of converted first-class constraints $\{O_I\}
\setminus \{G^i_0\}$ on appropriate Hilbert subspaces:
\begin{eqnarray}
\label{Q'} \hspace{-1em}&&\hspace{-1.0em}
 {Q}' = Q +
\eta^i_{G}(\sigma^i+h^i)+\mathcal{B}^i \mathcal{P}^i_{G},\quad
\texttt{with some $\mathcal{B}^i$,}\\
\label{Q} \hspace{-1em}&&\hspace{-0.5em}
 {Q} = \textstyle
 \Bigl(\frac{1}{2}q_0T_0 + \frac{1}{2}\eta_0L_0+ q_i^+T^i \eta_i^+L^i+\sum_{l\leq m}\eta_{lm}^+L^{lm}\nonumber\\
 \hspace{-1em}&&\hspace{-0em}
   + \textstyle\sum_{l<
m}\vartheta^+_{lm}T^{lm}+h.c.\Bigr)+ \frac{1}{2}
    \widehat{\mathcal{C}}^I\widehat{\mathcal{C}}^J{f}^K_{JI}\widehat{\mathcal{P}}_K,
 \\
\label{sigmai}
\hspace{-1em}&&\hspace{-0.5em}
   \sigma^i = G_0^i - h^i   - \textstyle\eta_i \mathcal{P}^+_i +
   \eta_i^+ \mathcal{P}_i +   q_ip_i^+ + q_i^+p_i \nonumber\\
   \hspace{-1em}&&\hspace{-0.5em}
   + \sum_{
m}(1+\delta_{im})(
\eta_{im}^+{\cal{}P}^{im}-\eta_{im}{\cal{}P}^+_{im}) \nonumber\\
   \hspace{-1em}&&\hspace{-0em} + \textstyle\sum_{l<i}[\vartheta^+_{li}
\lambda^{li} - \vartheta^{li}\lambda^+_{li}]-
\sum_{i<l}[\vartheta^+_{il} \lambda^{il} -
\vartheta^{il}\lambda^+_{il}]\,,
\end{eqnarray}
where $\{\widehat{\mathcal{C}}^I, \widehat{\mathcal{P}}^I\} \equiv
\{{\mathcal{C}}^I, {\mathcal{P}}^I\}\setminus \{\eta_{G}^i, \mathcal{P}_{G}^i\}$.~Next, we choose a representation of
$\mathcal{H}_{tot}$: $(q_i, p_i, \eta_i, \eta_{ij}$, $\vartheta_{rs},
\mathcal{P}_0$, $p_0, \mathcal{P}_i, \mathcal{P}_{ij}$, $\lambda_{rs},
\mathcal{P}^{i}_G)|0\rangle=0$, and suppose that the field vectors
$|\chi \rangle$ as well as the gauge parameters $|\Lambda \rangle$
do not depend on ghosts $\eta^{i}_G$:
\begin{eqnarray}\hspace{-1em}&&\hspace{-0.5em}
|\chi \rangle = \sum_n \prod_{l,i\le j, r<s}^k(f^+_l)^{n_l^0}( b_{ij}^+
)^{n_{ij}}( d_{rs}^+ )^{p_{rs}}q_0^{n_{b{}0}}\eta_0 ^{n_{f 0}}\nonumber
\\
\hspace{-1em}&&\hspace{-0.5em}\prod_{e, g,i, j,
l\le m, n\le o}\hspace{-2em}(q_e^+)^{n_{a{}e}}(p_g^+)^{n_{b{}g}}( \eta_i^+ )^{n_{f i}} ( \mathcal{P}_j^+ )^{n_{p
j}} ( \eta_{lm}^+ )^{n_{f lm}}
\nonumber
\\
\hspace{-1em}&&\hspace{-0.5em} ( \mathcal{P}_{no}^+ )^{n_{pno}}\prod\nolimits_{r<s, t<u}( \vartheta_{rs}^+)^{n_{f rs}}
( \lambda_{tu}^+ )^{n_{\lambda tu}}\times  \label{chi}
\\
\hspace{-1em}&&\hspace{-0.5em} |\Psi(a^+_i)^{n_{b{}0}\! n_{f 0};\!  (n)_{a{}e}\! (n)_{b{}g}\! (n)_{f
i}\!(n)_{p j}\!(n)_{f lm}\! (n)_{pno}\!(n)_{f rs}\!(n)_{\lambda
tu}}_{(n)^0_{l}(n)_{ij}(p)_{rs}}\rangle\footnotemark \nonumber.
\end{eqnarray}\footnotetext{The brackets $(n)_{f i},(n)_{p j}, (n)_{ij}$ in (\ref{chi}) means,
e.g., for $(n)_{ij}$ the set of indices $(n_{11},...$,
$n_{1k},..., n_{k1},..., n_{kk})$. The  sum above is taken over
$n_{b{}0}$, $n_{a{}e}$, $ n_{b{}g}$,   $n_{ij}$, $p_{rs}$ and  running from $0$ to infinity, and
over the rest $n$'s from $0$ to $1$.}We denote by $|\chi^k\rangle$
the state (\ref{chi}) satisfying to $gh(|\chi^k\rangle)=-k$. Thus,
the physical state having the ghost number zero is
$|\chi^0\rangle$, the gauge parameters $|\Lambda \rangle$ having
the ghost number $-1$ is $|\Lambda^0\rangle$ and so on. The vector
$|\chi^0\rangle$ must contain  physical string-like vector
$|\Psi\rangle = |\Psi(a^+_i)^{(0)_{f o} \ldots (0)_{rs}}\rangle$:
\begin{eqnarray}\label{decomptot}
\hspace{-1em}&&\hspace{-1.0em} |\chi^0\rangle = |\Psi\rangle+  |\Psi_A\rangle:\
|\Psi_A\rangle_{\big|\ [B^+_a = C^I = \mathcal{P}_I = 0]} =0.
\end{eqnarray}
Independence of the vectors (\ref{chi}) on $\eta^{i}_G$ transforms
the equation for the physical state ${Q}'|\chi^0\rangle = 0$ and the
BRST complex of the reducible gauge transformations,
$\delta|\chi\rangle$ = $Q'|\Lambda^0\rangle$, $\delta|\Lambda^0\rangle =
Q'|\Lambda^1\rangle$, $\ldots$, $\delta|\Lambda^{(r-1)}\rangle =
Q'|\Lambda^{(r)}\rangle$, to the relations:
\begin{eqnarray}
\label{Qchi} \hspace{-1em}&&\hspace{-0.5em}  \bigl(Q|\chi^0\rangle, \delta|\chi^0\rangle, ..., \delta|\Lambda^{(r-1)}\rangle\bigr) =\bigl(0, Q|\Lambda^0\rangle, ..., Q|\Lambda^{(r)}\rangle\bigr),\nonumber \\
\hspace{-1em}&&\hspace{-0.5em}  [\sigma^i+h^i]\bigl(|\chi^0\rangle, |\Lambda^0\rangle, \ldots ,
|\Lambda^{(r)}\rangle\bigr) = 0, \label{QLambda}
\end{eqnarray}
with  $r =
\sum_{l=1}^k n_l + k(k-1)/2 - 1$  being  the stage of reducibility both for
massless and for the massive fermionic HS field. Resolution  the
spectral  problem from the Eqs.(\ref{QLambda})
yields   the eigenvectors of
the operators $\sigma^i$: $|\chi^0\rangle_{(n)_k}$,
$|\Lambda^0\rangle_{(n)_k}$, $\ldots$, $|\Lambda^{r}\rangle_{(n)_k}$,
$n_1 \geq n_2 \geq \ldots n_k \geq 0$ and corresponding
eigenvalues of the parameters $h^i$ (for massless  HS fields and $i=1,..,k$),
\begin{eqnarray}
\label{hi} \hspace{-1em}&&\hspace{-1.5em}  -h^i = m_i+\textstyle\frac{d-4i}{2} \;, \
 m_1,...,m_{k-1} \in \mathbb{Z}, m_k \in
\mathbb{N}_0\,.
\end{eqnarray}
 One can show, first, the operator $Q$ is nilpotent on the subspaces determined by the solution for the Eq. (\ref{QLambda}), second,    to construct Lagrangian for the field corresponding to a definite YT
(\ref{Young k}) we must put  $m_i=n_i$, and, third,  one should substitute
$h^i$ corresponding to the chosen $n_i$ (\ref{hi})  into  $Q$
(\ref{Q'}) and relations (\ref{Qchi}).

To get the Lagrangian formulation with only first order derivatives, we, because of the functional dependence of the operator $L_0$ on fermionic one $T_0$, $L_0=-T_0^2$, may to gauge away a dependence on $L_0, \eta_0$ from the BRST operator $Q$ (\ref{Q}) and from the whole set of the vectors $|\chi^0\rangle_{(n)_k}$, $|\Lambda^{(s)}\rangle_{(n)_k}$. To do so, we extract the zero-mode ghosts from the operator
$Q$ as:
\begin{eqnarray}
\label{strQ}\hspace{-1em}&&\hspace{-0.5em}  Q = q_0\tilde{T}_0+\eta_0{L}_0 +
\imath(\eta_i^+q_i-\eta_iq_i^+)p_0\nonumber\\
\hspace{-1em}&&\hspace{-0em}  -
\imath(q_0^2-\eta_i^+\eta_i){\cal{}P}_0 +\Delta{}Q,
\end{eqnarray}
where the explicit form of $\Delta{}Q$ is easily restored from Eqs. (\ref{Q'}), (\ref{strQ}) and
\begin{eqnarray}
\label{tildeT0} \tilde{T}_0 &=& T_0 -2q_i^+{\cal{}P}_i
-2q_i{\cal{}P}_i^+:\ {\tilde{T}_0}^2 = - L_0.
\end{eqnarray}
 We also expand the state vector and
gauge parameters in powers of the zero-mode ghosts, for $s=0, \ldots, \sum_{o=1}^k n_o + k(k-1)/2-1$, $m=0,1$:
\begin{eqnarray}
\label{0chi} \hspace{-1em}&&\hspace{-1.0em} |\chi\rangle = \sum_{l\ge0}q_0^l(
|\chi_0^l\rangle +\eta_0|\chi_1^l\rangle),
gh(|\chi^{l}_{m}\rangle)=-(m+l),
\\
\label{0L} \hspace{-1em}&&\hspace{-1.0em}  |\Lambda^{(s)}\rangle
= \sum_{l=0}q_0^k(|\Lambda^{(s)}{}^l_0\rangle
+\eta_0|\Lambda^{(s)}{}^l_1\rangle).
\end{eqnarray}
Now, we may gauge away of all the fields and gauge parameters by means of the equations of motion and set of the gauge transformations (\ref{Qchi})
 except two,
$|\chi^0_0\rangle$, $|\chi^1_0\rangle$ for the fields and $|\Lambda^{(s)}{}^l_0\rangle$, for $l=0,1$ and $s=0, \ldots, r$, for the gauge parameters. To do so, we use in part the procedure described in \cite{symferm-flat}, \cite{mixfermiflat}.

As the result, the first-order equations of
motion  corresponding to the field with given spin
$(n_1+\frac{1}{2},...,n_k+\frac{1}{2})$ have the form in terms of the matrix notations,
\begin{eqnarray}
\hspace{-1em}&&\hspace{-0.5em} \left(\hspace{-0.4em}\begin{array}{cc}
\!\!\tilde{T}_0\! &\!\!\!\Delta{}Q\\
\!\!
\Delta{}Q \!& \!\!\!\frac{1}{2}
\bigl\{
   \tilde{T}_0,\eta_i^+\eta_i\bigr\}
   \end{array}\hspace{-0.6em}\right)\hspace{-0.3em}\left(\hspace{-0.5em}\begin{array}{c}
|\chi^{0}_{0}\rangle_(n)_k\\
|\chi^{1}_{0}\rangle_(n)_k
   \end{array}\hspace{-0.6em}\right)\hspace{-0.3em}  = \left(\hspace{-0.3em}\begin{array}{c}0\\0\end{array}\hspace{-0.3em}\right)\hspace{-0.3em}. \label{EofM1}
\end{eqnarray}
They are  Lagrangian
ones and can be deduced  from the  Lagrangian
action for fixed spin $(m)_k=(n)_k$, (being standardly defined up to an overall
factor and with omitting the subscript $(n)_k$)
\begin{eqnarray}
\hspace{-1em}&&\hspace{-1.2em}{\cal{}S}_{(n)_k}\hspace{-0.15em} =\hspace{-0.15em} \Bigl(\hspace{-0.1em}\langle\tilde{\chi}^{0}_{0}|, \langle\tilde{\chi}^{1}_{0}|\hspace{-0.1em}\Bigr)K\hspace{-0.4em}\left(\hspace{-0.4em}\begin{array}{cc}
\!\!\tilde{T}_0\! &\!\!\!\Delta{}Q\\
\!\!
\Delta{}Q \!& \!\!\!\frac{1}{2}
\bigl\{
   \tilde{T}_0,\eta_i^+\eta_i\bigr\}
   \end{array}\hspace{-0.6em}\right)\hspace{-0.3em}\left(\hspace{-0.5em}\begin{array}{c}
|\chi^{0}_{0}\rangle\\
|\chi^{1}_{0}\rangle
   \end{array}\hspace{-0.6em}\right)\hspace{-0.3em}
, \label{L1}
\end{eqnarray}
where the standard odd scalar product for the creation and
annihilation operators in $\mathcal{H}_{tot} = \mathcal{H}\otimes
\mathcal{H}' \otimes \mathcal{H}_{gh}$ is assumed and
non-degenerate  operator $K = K_{(n)_k}$ provides reality of the
action following from modifying Hermiticity for $o'_I$ in Section~\ref{Vermamodule}.
The action (\ref{L1})
is invariant with respect to the gauge
transformations, following from the tower of the Eqs. (\ref{Qchi}) with omitting $(n)_k$,
\begin{eqnarray}
\delta\left(\hspace{-0.5em}\begin{array}{c}
|\Lambda^{(s){}0}_{0}\rangle\\
|\Lambda^{(s){}1}_{0}\rangle
   \end{array}\hspace{-0.6em}\right)\hspace{-0.3em} = \hspace{-0.3em}\left(\hspace{-0.4em}\begin{array}{cc}
\!\!
\Delta{}Q \!& \!\!\!\frac{1}{2}
\bigl\{
   \tilde{T}_0,\eta_i^+\eta_i\bigr\} \\
 \!\!\tilde{T}_0\! &\!\!\!\Delta{}Q
   \end{array}\hspace{-0.6em}\right)\hspace{-0.3em} \left(\hspace{-0.5em}\begin{array}{c}
|\Lambda^{(s+1){}0}_{0}\rangle\\
|\Lambda^{(s+1){}1}_{0}\rangle
   \end{array}\hspace{-0.6em}\right)\hspace{-0.3em},
\label{GT1}
\end{eqnarray}
for $s=-1,0,\ldots, \sum_{o=1}^k n_o + k(k-1)/2-1$, and $|\Lambda^{(-1){}l}_{0}\rangle \equiv |\chi^{l}_{0}\rangle$.

Concluding, one can prove  the  action (\ref{L1}) indeed
reproduces the basic conditions (\ref{Eq-1})--(\ref{Eq-3}) for
massless  (massive) HS fields. General action (\ref{L1}) gives,
in principle, a straight recept to obtain the Lagrangian for any
component field from general vectors $| \chi^l_0 \rangle_{(n)_k}$.
\section{Conclusion}

Thus, we have constructed a gauge-invariant unconstrained
Lagrangian description of free half-integer  HS fields belonging to an
irreducible representation of the Poincare group $ISO(1,d-1)$ with
the arbitrary Young tableaux having $k$ rows in the ``metric-like"
formulation. The results of this study are the general and
obtained on the base of universal method which is applied by the
unique way to both massive and massless bosonic HS fields with a
mixed symmetry in a Minkowski space of any dimension.


\section*{Acknowledgement}
The author is grateful to E. Skvortsov, K. Stepanyantz, V. Krykhtin, D. Francia  for valuable comments and to I.L. Buchbinder for
collaboration on many stages of the work. The work was supported  by
 the RFBR grant, project  Nr. 12-02-00121
and by LRSS grant Nr.224.2012.2.

\end{multicols}
\end{document}